\documentclass[structabstract]{aa}  
\usepackage{natbib}
\bibpunct{(}{)}{;}{a}{}{,} 

\usepackage{graphicx}
\usepackage{txfonts}
\begin{document}

\title{Line-profile variations in radial-velocity measurements}

    \subtitle{Two alternative indicators for planetary searches}

\author{P. Figueira\inst{1}
             \and 
        N. C. Santos\inst{1,2}
	     \and
         F. Pepe\inst{3}
         \and 
         C. Lovis\inst{3}
	\and N. Nardetto\inst{4}
         }

   \institute{Centro de Astrof\'{i}sica, Universidade do Porto, Rua das Estrelas, 4150-762 Porto, Portugal\\
     \email{pedro.figueira@astro.up.pt}
	\and
	Departamento de F\'{i}sica e Astronomia, Faculdade de Ci\^{e}ncias, Universidade do Porto, Portugal 
	\and
	 Observatoire Astronomique de l'Universit\'{e} de Gen\`{e}ve, 51 Ch. des Maillettes, 
    - Sauverny - CH1290, Versoix, Suisse 
	\and
	Laboratoire Lagrange, UMR7293, UNS/CNRS/OCA, 06300 Nice, France 
	}

   \date{}

 
  \abstract
   {}
   {We introduce two methods to identify false-positive planetary signals in the context of radial-velocity exoplanet searches. The first is the bi-Gaussian cross-correlation function fitting (and monitoring of the parameters derived from it), and the second is the measurement of asymmetry in radial-velocity spectral line information content, $V_{asy}$. We assess the usefulness of each of these methods by comparing their results with those delivered by current indicators.}
   {We make a systematic analysis of the most used common line profile diagnosis, Bisector Inverse Slope and Velocity Span, along with the two proposed ones. We evaluate all these diagnosis methods following a set of well-defined common criteria and using both simulated and real data. We apply them to simulated cross-correlation functions that are created with the program SOAP and which are affected by the presence of stellar spots. We consider different spot properties on stars with different rotation profiles and simulate observations as obtained with high-resolution spectrographs. We then apply our methodology to real cross-correlation functions, which are computed from HARPS spectra, for stars with a signal originating in activity (thus spots) and for those with a signal rooted on a planet.}
   {We demonstrate that the bi-Gaussian method allows a more precise characterization of the deformation of line profiles than the standard bisector inverse slope. The calculation of the deformation indicator is simpler and its interpretation more straightforward. More importantly, its amplitude can be up to 30\% larger than that of the bisector span, allowing the detection of smaller-amplitude correlations with radial-velocity variations. However, a particular parametrization of the bisector inverse slope is shown to be more efficient on high-signal-to-noise data than both the standard bisector and the bi-Gaussian. The results of the $V_{asy}$ method  show that this indicator is more effective than any of the previous ones, being correlated with the radial-velocity with more significance for signals resulting from a line deformation. Moreover, it provides a qualitative advantage over the bisector, showing significant correlations with RV for active stars for which bisector analysis is inconclusive.}
   {We show that the two indicators discussed here should be considered as standard tests to check for the planetary nature of a radial-velocity signal. We encourage the usage of different diagnosis as a way of characterizing the often elusive line profile deformations.}

  \keywords{(Stars:) Planetary systems, Techniques: radial velocities, Line: profiles, Methods: data analysis }

\authorrunning{P. Figueira et al.}
\titlerunning{Line profile indicators for planetary searches}

   \maketitle
%

\section{Introduction}

The detection of extrasolar planets relies heavily on the radial velocity (RV) method, which accounts for the vast majority of the planetary discoveries up to now. Yet, it has been known since the very first announcement by \cite{1995Natur.378..355M} that this approach is vulnerable to false positives. The potential parasitic signals are often rooted in stellar photospheric effects that deform spectral line profiles and introduce a RV signal which is not rooted on a Doppler shift of the spectral lines.

The relatively high frequency of false positives led to the establishment of studies of line profile properties for detected RV signals as a way of proving -- or better said, disproving -- their planetary nature. The first in-depth study was done by \cite{2001A&A...379..279Q}, whom established the bisector and the bisector inverse slope ($BIS$) as the paradigm for its genre. The authors showed that there was a clear correlation between $BIS$ and the measured RV for the case of a signal created by a photospheric spot. Numerous studies have made use of this principle \citep[e.g.][just to cite a few]{2005A&A...442..775M, 2009ApJ...707..768N, 2010A&A...512A..46H, 2010A&A...513L...8F} to reject the planetary nature of a signal, and planet-hunting surveys now routinely search for a correlation between RV and $BIS$ variations as a way of pinpointing a parasite signal.

Several alternative methodologies exist for identifying these signals. For stabilized spectrographs, the monitoring of the {\it FWHM} of the cross-correlation function can reveal profile variations \citep[e.g.][]{2011A&A...535A..55D}, and the $V_{span}$ was introduced by \cite{2011A&A...528A...4B} as an alternative to BIS to handle cases of low-S/N data. Moreover, several campaigns compared RV measurements made in different wavelengths to exclude planetary candidates, since a cold spot will induce a signal whose amplitude depends on the contrast relative to the surrounding stellar disk \citep[e.g.][]{2006ApJ...644L..75M, 2008A&A...489L...9H, 2010A&A...511A..55F}. When the range of possible applications is wider, an indicator is more used, and the bisector, making very few assumptions and being easy to calculate, is by far the most popular.  

In a general sense, the efficiency of a diagnosis criteria in identifying RV signals created by line deformations depends mostly on two properties. First, the indicator should be proportional to the RV signal created (if possible linearly proportional, as it eases its identification and interpretation), and second this proportionality should be characterized by a proportionality constant as large as possible (when compared with the average error bars of the indicator).

Even though $BIS$ is undoubtedly useful in many situations, studies such as those of \cite{1997ApJ...485..319S}, \cite{2003A&A...406..373S}, and \cite{2007A&A...473..983D} pointed to an important limitation. For many cases of line profile deformations, characterized by certain stellar and spot properties, the $BIS$ is intrinsically less sensitive to the presence of a deformation than the RV induced by the deformation itself. In other words, the proportionality constant between $BIS$ and RV for these configurations has an absolute value $\ll$\,1. On top of this, it is well known that the relation between $BIS$ and RV is not expected to be monotonic \citep[see e.g.][]{2011A&A...528A...4B}. While for many practical cases the correlation can be assumed to be a straight line, the real shape is similar to that of a tilted ``8''; the proportionality constant between the two variables can even change from positive to negative or vice-versa for some cases. These two properties can preclude the detection of a correlation between $BIS$ and $RV$ and render the diagnostic ineffective.

As we expand our searches, including more varied stellar hosts and pushing the precision of our surveys further to detect lower-amplitude signals, there is an increasing demand for alternative and more effective ways of identifying false-positive signals. There have been few attempts at finding alternative methods of characterizing line profile variations in a way useful for planetary RV searches \citep[with the work of][being a rare exception]{2006A&A...454..341D}.

In this paper, we consider two different diagnosis methods and compare their performance with those of current line-profile indicators for a wide variety of cases. We start by presenting the definition and properties of $BIS$ and $V_{span}$, our comparison references, in Sect.\,2. Then we describe the data sets used for this study and the principles of our comparison in Sect.\,3. We employ the standard line-profile indicators and evaluate their results in Sect.\,4. We move on to apply the first of our two methods, the Bi-Gaussian fitting, in Sect.\,5 and the second, $V_{asy}$, in Sect.\,6. We discuss the results in Sect.\,7 and conclude on the usefulness of these new indicators and how they help us understand the impact of line-profile variation on RV in Sect.\,8.


\section{Current diagnosis methods}

The spectral information content of a stellar spectrum can be condensed in a cross-correlation function (CCF), which is then used to measure precise RVs \citep{1996A&AS..119..373B, 2002A&A...388..632P}. The $RV$ is extracted by fitting a Gaussian function to the CCF with the center of the Gaussian delivering the star's RV; this quantity is represented throughout the paper as $RV_{G, center}$. We apply our methods to the CCF and its derived quantities and compare the results with those obtained for known line-profile indicator variations. For reference line-profile indicators, we consider the $BIS$, the most common diagnosis method for the presence of parasitic signals, and $V_{span}$. In the next sections we describe these methods in detail and how they were implemented.

\subsection{The $BIS$}

In the $BIS$ analysis, we used the definition provided by \cite{2001A&A...379..279Q}. Consider a CCF and calculate its mid-point (i.e., the point at equal distance between its blue and red wing) for each flux level between the continuum level and the bottom of the line. Since the flux level will only coincidentally fall on a pixel value, the flux values are calculated by interpolation from the neighboring CCF pixels.  The $BIS$ value is then simply the difference between the velocity of the top and bottom sections of the line, with the top being the average mid-point of the flux levels located between 60 and 90\% and the bottom of those between 10 and 40\% of the line full depth. We refer the reader to the Fig.\,5 of \cite{2001A&A...379..279Q} for a helpful schematic representation. Since we are working in the velocity space, the line mid points are measured in km/s (or any multiple of it), as is $BIS$.

The interpretation of the $BIS$ is quite straightforward. Any $RV$ signal associated with a variation between the top and bottom sections of the line is not rooted in the coordinated movement of it but rather on a line deformation, and a plot of $BIS$-$RV$ will reveal a correlation. 

Even though the vast majority of works follow this prescription (see references in the introduction), the top and bottom regions of the $BIS$ can be defined differently. In particular, one can consider narrower bands located at different heights, leading to different slopes when the line is subjected to deformations. This different leverage capability will lead to different indicator amplitude variations at the expense of using only a smaller section of the line, and thus working at a lower signal-to-noise ratio when applying it to real spectra. 
To explore this posibility, we considered two different parametrizations, representative of extreme cases: a maximum leverage case $BIS^+$ in which the limits of top and bottom regions are 80-90\% and  10-20\% of the total flux level, respectively, and a minimum leverage case $BIS^-$ with top at 60-70\% and bottom at 30-40\%, respectively.

\subsection{The $V_{span}$}

We employ the $V_{span}$ indicator as defined in \cite{2011A&A...528A...4B}. This indicator represents the difference $RV_{high} - RV_{low}$ between two Gaussian fits of the CCF which consider exclusively the upper and lower part of the CCF, respectively. 
For the first, $RV_{high}$, one takes into account only the top of the CCF for the fit by considering only the points in the range [$-\infty$:$-\sigma$][$+\sigma$:$+\infty$] as measured relative to the center of the Gaussian fit; for the second $RV_{low}$ one considers the points in the range [$–\infty$:$-3\sigma$]$\vee$[$–\sigma$:$+\sigma$]$\vee$[$+3\sigma$:$+\infty$]. The former quantity is sensitive to variations on the top of the line and the latter on the bottom. 

The construction of the $V_{span}$ indicator was motivated by the analysis of line-profile variations for cases of low S/N and was shown to be advantageous relative to the $BIS$ for one of these cases. We refer the interested reader to \cite{2011A&A...528A...4B} for details.

\section{Choice of datasets \& comparison principles}

\subsection{The datasets}

We simulated CCFs as affected by the presence of a stellar spot using the program SOAP, and refer the interested reader to \cite{2012A&A...545A.109B} for details on the program. All simulations share two basic assumptions: we acquire our spectra with a spectrograph with a resolution of 100 000 (which defines the {\it FWHM} of a non-rotating star), and we observe a star with a single dark spot (i.e, with zero emissivity). We consider different {\bf $v\,\sin{i}$} for the star, different inclinations of the stellar rotation axis relative to the line of sight, different spot latitudes, and different spot filling factors. When performing our simulations, we use a grid resolution of 2000 and a spot circumference resolution of 50; the orbital phase is homogeneously sampled by 100 points. For the steps in RV for the CCF calculation, we considered a very fine grid with steps of 10\,m/s from -30\,km/s to +30\,km/s. On real data the CCF is calculated on a much coarser grid, usually of $\sim$100\,m/s or higher. However, a coarse discretization can introduce an appreciable deviation in the fit and on the fit parameters. Since these synthetic data have, by definition, no noise, any fit instability steming from the small number of points can introduce a detectable difference relative to high S/N spectra with a coarser grid. Since computational power is not an issue for such a small number of simulations, we choose to calculate the CCF on a very fine grid to evaluate the application of our indicators in the most meaningfull way possible.

We started by simulating the properties of standard well-studied cases, with the first being that of HD\,166435, already discussed both in \cite{2001A&A...379..279Q} and in \cite{2007A&A...473..983D}. The star has a $v\sin{i}$\,=\,7.0\,km/s and is observed at an inclination of 30\,$^o$; its spot has a filling factor of 1\% and is located at a latitude of 60\,$^o$. This is the most extensively studied spot configuration in the literature, and we included it here for that reason\footnote{ We note that the spot latitude definition used in {\it SOAP} is not the same as that used by \cite{2007A&A...473..983D}; in the former, the latitude angle is used, while the colatitude angle is used in the latter.}.

We then moved to more general cases with the objective of studying the impact of key stellar and spot parameters on the amplitude of the measured RVs and the associated indicator's response. First we studied the impact of equatorial spots (latitude of 0\,$^o$) with different filling factor values of 1.0, 2.0, 5.0, and 10.0 \% on a edge-on star (I\,=\,0$^o$) with $v\sin{i}$\,=\,7.0\,km/s. Then, we repeated the exercise for a 1\% filling factor equatorial spot on a edge-on star with rotational velocities of 2, 5, and 10\,km/s. These tests were also inspired from those of \cite{2007A&A...473..983D}, who chose these inclination and spot latitudes to induce the highest $RV$ amplitude for a given filling factor and $v\sin{i}$. The spot filling factors were chosen to depict typical spot sizes on main-sequence stars, and the $v\sin{i}$ were chosen to illustrate several cases ranging from a broadening inferior to that introduced by the resolution of the spectrograph, to a star rotating so fast that the RV precision is significantly degradated.

We finally tested our approach on real data, and to do so we used HARPS spectra and CCFs as delivered by the last version of the DRS pipeline \citep{2003Msngr.114...20M}. We selected both active stars  (log($R'_{HK}$) $\sim$ -4.46 or higher), for which a correlation between $RV$ and the indicators ($\Delta V$ and {\it BIS}) is expected, and a planet-host star HD\,216770, for which it is not (or is not expected at the level of the measured $RV$ signal amplitude). The planetary orbits measured on this star were discussed in \cite{2004A&A...415..391M}, and the reader is referred to the paper for details. The stars are listed in Table\,\ref{stars_act}, along with the log($R'_{HK}$), $v\sin{i}$, and the number of data points taken at the time of the publication and considered for the analysis. The $v\,\sin{i}$ was derived in \cite{2003A&A...398..363S} and log($R'_{HK}$) was derived in \cite{2000A&A...361..265S}. The only exception was HD\,216770, of which log($R'_{HK}$) was taken from \cite{1996AJ....111..439H}.


\begin{table}

\caption{HARPS star properties:} \label{stars_act}

\centering
\begin{tabular}{lcccc} \hline\hline
 \ \ star & $R'_{\mathrm{HK}}$ & $v.\sin{i}$ (km/s) & \# points  &  comments \\ 
\hline 

HD\,224789 &  -4.46 & 3.01 & 36  & active star\\
HD\,36051 & -4.55 & 5.07 & 13  & active star \\
HD\,103720 & -4.46 & 2.82 & 49  & active star \\
HD\,200143 & -4.48 & 3.02 & 42  & active star \\
BD-213153 & -4.51 & 3.63 & 29  & active star \\
\hline
HD\,216770 & -4.84 & 1.4 & 29 & planet-host \\
\hline
\end{tabular}

\end{table}

When the activity of the star is larger, the impact on both measured RV and line-profile variation indicators is larger. To maximize its impact on RV, one should measure it at least over one full activity cycle. In practice, this is very difficult, and recent works have shown that several longer-term activity cycles are present in a star on top of the typical rotation-modulated cycle, with a strong impact on RV \citep[e.g.][]{2011A&A...535A..55D, 2011arXiv1107.5325L}. While keeping in mind that the characterization of the activity-induced effects is still a work in progress, we concentrate our analysis on the {\it relative} variations between RV and line-profile indicators and how the different indicators compare with each other.

\subsection{Comparing the datasets}

We aim to compare the different indicators in the most systematic way possible. To do so, we apply them to the same datasets and evaluate them according to the same criteria.

We first compare the amplitude of the indicator's variation (relative to its average value) with those of the $RV$ and other indicators. We pay particular attention to how the indicators perform compared to $BIS$, the most commonly used line-profile indicator. If one assumes that the error bars are similar, a larger amplitude of the indicator's variation results in a greater capability to pinpoint a signal. Yet, the indicator's capabilities must be assessed through a rigorous quantitative analysis. To compare the correlation of an indicator with $RV$, we also consider:

\begin{itemize}
 
 \item the slope resulting from the fit of a linear function by least-square minimization ($m_{\mathrm{ind}}$);
 \item the Pearson's correlation coefficient value for the data sets ($\rho_\mathrm{{ind}}$); and
 \item the probability of obtaining an equal or larger correlation coefficient absolute value by chance.
\end{itemize}

The slope of the correlation between an indicator and the $RV$ is a different way of relating a variation in the indicator with that of the RV. However, it does not evaluate the strength of the correlation betwen the two data sets. To do this, we use the Pearson's correlation coefficient, which is a measure of linear dependence between the two variables. To test if the correlation coefficient's value is significant, we test if it can be obtained by a fortuitous pair-matching between the indicator and the RV. To do so, we bootstrap 100 000 times the data pairs and calculate the correlation coefficient $\rho_{\mathrm{ind}}$ for each of these synthetic uncorrelated data sets. We then calculate the average and standard deviation ($\sigma_\mathrm{{\rho, ind}}$) of these $\rho_{\mathrm{ind}}$ values; by assuming a Gaussian distribution, one can calculate the probability that the $\rho_{\mathrm{ind}}$ of the original dataset was obtained by pure chance. While the probability value is conditioned by the hypothesis that the indicators follow a Gaussian distribution, its distance in $\sigma$ is still a good indicator on how the measured value compares with that of an uncorrelated distribution.

When using real data, we have two important additional information elements. The first of these is the $RV$ measurement uncertainty, as estimated from the $RV$ information content of the spectra \citep[e.g.][]{2001A&A...374..733B}. Moreover, we know if the $RV$ signal introduced is of planetary or stellar origin and, as a consequence, if a line profile deformation is present or not; we also know then if the indicator is expected to be correlated with the measured $RV$ variation or not. It follows then that we can add the following criteria when evaluating the real data sets:

\begin{itemize}
 \item the error bars on the correlation slope $(m)_{\mathrm{ind}}$ which result from bootstrapping the $RV$ data and indicator with respect to the respective error bars, performed by drawing 100 000 times values around their error bars;
 \item the relationship of the slope and correlation values to the stellar activity or presence of planets around the star.
 \end{itemize}

We note that the bootstrapping is a non-parametric test, depending on very few assumptions, and thus being very robust. However, to take full advantage of the real data, one has to estimate the error bars on the indicators themselves. In spite of the abundant literature on $BIS$, there are  virtually no studies on the error distribution of this estimator. It is commonly reported that the error bar on this difference is on the order of $\sqrt{2}$ times the photon noise on the CCF because $BIS$ corresponds to a difference of velocities between the upper and the lower sections of the CCF. Even though these error bars are ill-justified, our objective is to compare the performance among indicators, and and so we choose to use the commonly accepted uncertainties for $BIS$ for the other indicators. To allow a more straightforward interpretation of our results, we separate those that depend on the assumptions on the error bars from those which do not and present them in different tables.


\section{Standard profile indicators: $BIS$ and $V_{span}$}

\subsection{$BIS$ {\it vs} $V_{span}$}

\begin{table*}

\caption{Comparison between correlations of $BIS$ and $V_{span}$ with $RV$ for simulated data.} \label{Vspan_simul}

\centering
\begin{tabular}{cccccccccc} \hline\hline
 \ \ model\,/\,star & $RV_{max}$ & $BIS_{max}$ & $V_{span, max}$  & $m_{BIS}$ & $m_{V_{span}}$ & $\rho_{BIS}$ & $\rho_{V_{span}}$ & $\sigma_{\rho, BIS}$ & $\sigma_{\rho, V_{span}}$ \\ 
\hline 

HD\,166435' & 9.28e+01	 & 	1.03e+02 	 & 	9.04e+01	 & 	-1.068	 & 	-0.917	 & 	-0.990	 & 	-0.983 &  9.86 &  9.73 \\

\hline  

$F_r$\,=\,1.0\% & 9.98e+01	 & 	1.03e+02 	 & 	8.95e+01	 & 	-0.686	 & 	-0.569	 & 	-0.827	 & 	-0.802 &  8.23 &  7.99 \\
$F_r$\,=\,2.0\% & 2.01e+02	 & 	2.10e+02 	 & 	1.82e+02	 & 	-0.679	 & 	-0.563	 & 	-0.820	 & 	-0.793 &  8.18 &  7.88 \\
$F_r$\,=\,5.0\% & 5.20e+02	 & 	5.50e+02 	 & 	4.87e+02	 & 	-0.655	 & 	-0.540	 & 	-0.795	 & 	-0.760 &  7.90 &  7.58 \\
$F_r$\,=\,10.0\% & 1.10e+03	 & 	1.22e+03 	 & 	1.10e+03	 & 	-0.599	 & 	-0.493	 & 	-0.741	 & 	-0.693 &  7.37 &  6.90 \\

\hline  

$v.\sin{i}$=2.0  & 2.21e+01	 & 	1.94e+00 	 & 	1.42e+00	 & 	-0.057	 & 	-0.042	 & 	-0.801	 & 	-0.800 &  7.97 &  7.94 \\
$v.\sin{i}$=5.0  & 6.43e+01	 & 	3.72e+01 	 & 	2.95e+01	 & 	-0.378	 & 	-0.293	 & 	-0.815	 & 	-0.805 &  8.09 &  8.02 \\
$v.\sin{i}$=10.0  & 1.57e+02	 & 	2.54e+02 	 & 	2.53e+02	 & 	-1.109	 & 	-1.001	 & 	-0.847	 & 	-0.793 &  8.43 &  7.86 \\

\hline

\end{tabular}

\tablefoot{When not explicitly stated, the probability of obtaining the presented $\rho$ values from the distribution obtained by bootstrapping is lower than 1.0e-4. $RV$, $BIS$ and $V_{span}$ in units of m/s.}

\end{table*}

\begin{table*}

\caption{Comparison between $BIS$ and  $V_{span}$  correlations with $RV$ for HARPS data for the parameters that do not take into consideration the error bars.} \label{Vspan_real}

\centering
\begin{tabular}{cccccccc} \hline\hline
 \ \ star & $RV_{max}$ & $BIS_{max}$ & $V_{span}$ & $\rho_{BIS}$ & $\rho_{V_{span}}$ & $\sigma_{\rho, BIS}$ & $\sigma_{\rho, V_{span}}$ \\ 
\hline 

HD\,224789 &   3.53e+01	 & 	2.77e+01 	 & 	2.10e+01	 &  	-0.818	 & 	-0.818 &  4.84 ($<$0.00 \%) &  4.84 ($<$0.00 \%) \\

HD\,36051 &  2.16e+01	 & 	3.29e+01 	 & 	2.03e+01	 &  	-0.275	 & 	-0.286 &  0.95 (34.11 \%) &  0.99 (32.09 \%) \\

HD\,103720 &  1.10e+02	 & 	2.41e+01 	 & 	1.87e+01	 &  	-0.243	 & 	-0.314 &  1.68 ( 9.24 \%) &  2.17 ( 2.99 \%) \\

HD\,200143 &  4.70e+01	 & 	1.78e+02 	 & 	1.53e+02	 &  	-0.302	 & 	-0.260 &  1.93 ( 5.35 \%) &  1.66 ( 9.61 \%) \\

BD-213253 &  5.33e+01	 & 	3.33e+01 	 & 	2.45e+01	 &  	-0.130	 & 	-0.135 &  0.69 (49.18 \%) &  0.71 (47.70 \%) \\

\hline  

HD\,216770 & 2.67e+01	 & 	4.95e+00 	 & 	4.30e+00	 &  	0.142	 & 	0.178 &  0.75 (45.11 \%) &  0.94 (34.85 \%) \\

\hline

\end{tabular}

\tablefoot{$RV$, $BIS$ and $V_{span}$ in m/s.}

\end{table*}

\begin{table*}

\caption{Comparison between $BIS$ and $V_{span}$ correlations with $RV$ for HARPS data for the parameters that consider the error bars, i.e. the slope fit parameters.} \label{Vspan_real_MC}

\centering
\begin{tabular}{ccc} \hline\hline
 \ \ star & $(m,b)_{BIS}$ & $(m,b)_{V_{span}}$  \\ 
\hline 
\noalign{\vskip0.01\columnwidth} 

HD\,224789 &  (  -0.705 $_{-  0.011}^{+  0.012}$   or $_{-  1.57\%}^{+  1.73\%}$,     26.083 $_{-  0.451}^{+  0.409}$ ) &  (  -0.525 $_{-  0.010}^{+  0.011}$   or $_{-  1.99\%}^{+  2.13\%}$,     19.424 $_{-  0.413}^{+  0.386}$ )  \\[4pt]

HD\,36051 &  (  -0.476 $_{-  0.126}^{+  0.144}$   or $_{- 26.51\%}^{+ 30.16\%}$,      0.795 $_{-  0.232}^{+  0.204}$ ) &  (  -0.309 $_{-  0.112}^{+  0.124}$   or $_{- 36.46\%}^{+ 40.27\%}$,      0.517 $_{-  0.200}^{+  0.182}$ ) \\[4pt]

HD\,103720 &  (  -0.049 $_{-  0.005}^{+  0.006}$   or $_{- 11.26\%}^{+ 11.39\%}$,     -0.287 $_{-  0.036}^{+  0.036}$ ) &  (  -0.045 $_{-  0.005}^{+  0.006}$   or $_{- 12.02\%}^{+ 12.14\%}$,     -0.276 $_{-  0.036}^{+  0.036}$ )  \\[4pt]

HD\,200143 &  (  -0.239 $_{-  0.021}^{+  0.022}$   or $_{-  8.61\%}^{+  9.26\%}$,     -4.887 $_{-  0.421}^{+  0.452}$ ) &  (  -0.158 $_{-  0.020}^{+  0.021}$   or $_{- 12.71\%}^{+ 13.40\%}$,     -3.233 $_{-  0.410}^{+  0.433}$ ) \\[4pt]

BD-213253 &  (  -0.225 $_{-  0.032}^{+  0.034}$   or $_{- 14.09\%}^{+ 15.09\%}$,      5.527 $_{-  0.829}^{+  0.774}$ ) &  (  -0.172 $_{-  0.030}^{+  0.032}$   or $_{- 17.67\%}^{+ 18.53\%}$,      4.220 $_{-  0.778}^{+  0.742}$ ) \\[2pt]

\hline 
\noalign{\vskip0.01\columnwidth} 

HD\,216770 &  (   0.014 $_{-  0.006}^{+  0.006}$   or $_{- 42.97\%}^{+ 43.20\%}$,     -0.439 $_{-  0.194}^{+  0.193}$ ) &  (   0.013 $_{-  0.006}^{+  0.006}$   or $_{- 47.49\%}^{+ 47.62\%}$,     -0.401 $_{-  0.193}^{+  0.192}$ ) \\[2pt]

\hline

\end{tabular}
\end{table*} 

We start by applying the two standard line-profile indicators to our datasets. These have been discussed in their presentation papers and an exhaustive analysis is not the objective of this paper, so we will focus intead on their relative strenghts and weaknesses.

Both the analysis of synthetic (Table\,\ref{Vspan_simul}) and real data (Table\,\ref{Vspan_real} and Table\,\ref{Vspan_real_MC}) show that the $BIS$ and $V_{span}$ yields similar results for most of the cases. Their variation is proportional to both the spot filling factor and the star rotational velocity. For most of the considered cases, the results are of the same order of magnitude for the data and the different evaluation criteria considered. $BIS$ exhibits higher-amplitude variations and more significant correlations in a consistent way for all synthetic datasets but does not prove to be advantageous when applied to real data. $V_{span}$ yielded poorer results than those delivered by the $BIS$ on what concerns correlation slopes and associated errors, but we must stress that the latter depends on the assumed error bars for the bootstrapping.

\subsection{Different parametrizations of $BIS$}

\begin{table*}

\caption{Comparison between correlations of $BIS^+$ and $BIS^-$ with $RV$ for simulated data.} \label{BISES_simul}

\centering
\begin{tabular}{cccccccccc} \hline\hline
 \ \ star & $RV_{max}$ & $BIS^+_{max}$ & $BIS^-_{max}$ & $m_{BIS^+}$ & $m_{BIS^-}$ & $\rho_{BIS^+}$ & $\rho_{BIS^-}$ & $\sigma_{\rho, BIS^+}$ & $\sigma_{\rho, BIS^-}$ \\ 
\hline 

HD\,166435' & 9.28e+01	 & 	1.50e+02 	 & 	5.69e+01	 & 	-1.548	 & 	-0.593 	 & 	-0.988	 & 	-0.993 &  9.85 &  9.90 \\

\hline  

$F_r$\,=\,1.0\% & 9.98e+01	 & 	1.50e+02 	 & 	5.72e+01	 & 	-0.993	 & 	-0.383 	 & 	-0.824	 & 	-0.833 &  8.20 &  8.27 \\
$F_r$\,=\,2.0\% & 2.01e+02	 & 	3.03e+02 	 & 	1.17e+02	 & 	-0.982	 & 	-0.380 	 & 	-0.816	 & 	-0.827 &  8.15 &  8.22 \\
$F_r$\,=\,5.0\% & 5.20e+02	 & 	7.92e+02 	 & 	3.19e+02	 & 	-0.941	 & 	-0.370 	 & 	-0.788	 & 	-0.804 &  7.86 &  8.01 \\
$F_r$\,=\,10.0\% & 1.10e+03	 & 	1.68e+03 	 & 	7.79e+02	 & 	-0.849	 & 	-0.346 	 & 	-0.727	 & 	-0.745 &  7.20 &  7.41 \\

\hline 

$v.\sin{i}$=2.0  & 2.21e+01	 & 	2.89e+00 	 & 	1.05e+00	 & 	-0.085	 & 	-0.031 	 & 	-0.801	 & 	-0.799 &  7.98 &  7.95 \\
$v.\sin{i}$=5.0  & 6.43e+01	 & 	5.29e+01 	 & 	2.15e+01	 & 	-0.541	 & 	-0.216 	 & 	-0.818	 & 	-0.811 &  8.17 &  8.06 \\
$v.\sin{i}$=10.0  & 1.57e+02	 & 	4.10e+02 	 & 	1.26e+02	 & 	-1.699	 & 	-0.557 	 & 	-0.818	 & 	-0.879 &  8.17 &  8.74 \\

\hline

\end{tabular}

\tablefoot{When not explicitly stated, the probability of obtaining the presented $\rho$ values from the distribution obtained by bootstrapping is lower than 1.0e-4. $RV$, $BIS^+$, and $BIS^-$ in units of m/s.}

\end{table*}

\begin{table*}

\caption{Comparison between $BIS^+$ and $BIS^-$ correlations with $RV$ for HARPS data for the parameters that do not consider the error bars.} \label{BISES_real}

\centering
\begin{tabular}{cccccccc} \hline\hline
 \ \ star & $RV_{max}$ & $BIS^+_{max}$ & $BIS^-_{max}$ & $\rho_{BIS^+}$ & $\rho_{BIS^-}$ & $\sigma_{\rho, BIS^+}$ & $\sigma_{\rho, BIS^-}$ \\ 
\hline 

HD\,224789 &  3.53e+01	 & 	4.15e+01 	 & 	1.51e+01	 &  	-0.819	 & 	-0.795 &  4.84 ($<$0.00 \%) &  4.71 ($<$0.00 \%) \\

HD\,36051 &  2.16e+01	 & 	7.83e+01 	 & 	3.37e+01	 &  	-0.459	 & 	0.116 &  1.59 (11.16 \%) &  0.40 (68.64 \%) \\

HD\,103720 &  1.10e+02	 & 	3.86e+01 	 & 	1.55e+01	 &  	-0.214	 & 	-0.290 &  1.49 (13.66 \%) &  2.00 ( 4.49 \%) \\

HD\,200143 &  4.70e+01	 & 	1.78e+02 	 & 	4.45e+01	 &  	-0.343	 & 	-0.379 &  2.20 ( 2.82 \%) &  2.44 ( 1.51 \%) \\

BD-213253 &  5.33e+01	 & 	6.96e+01 	 & 	2.01e+01	 &  	0.031	 & 	-0.307 &  0.16 (87.10 \%) &  1.62 (10.46 \%) \\

\hline  

HD\,216770 &   2.67e+01	 & 	6.33e+00 	 & 	4.83e+00	 &  	0.127	 & 	0.137 &  0.67 (50.09 \%) &  0.73 (46.83 \%) \\

\hline

\end{tabular}

\tablefoot{$RV$, $BIS^+$ and $BIS^-$ in m/s.}

\end{table*}

\begin{table*}

\caption{Comparison between $BIS^+$ and $BIS^-$ correlations with $RV$ for HARPS data for the parameters that consider the error bars, i.e. the slope fit parameters.} \label{BISES_real_MC}

\centering
\begin{tabular}{ccc} \hline\hline
 \ \ star & $(m,b)_{BIS^+}$ & $(m,b)_{BIS^-}$  \\ 
\hline 
\noalign{\vskip0.01\columnwidth} 

HD\,224789 &  (  -1.018 $_{-  0.013}^{+  0.014}$   or $_{-  1.24\%}^{+  1.38\%}$,     37.647 $_{-  0.519}^{+  0.466}$ ) &  (  -0.379 $_{-  0.010}^{+  0.011}$   or $_{-  2.69\%}^{+  2.83\%}$,     14.014 $_{-  0.395}^{+  0.376}$ ) \\[4pt]

HD\,36051 &  (  -1.164 $_{-  0.163}^{+  0.209}$   or $_{- 14.01\%}^{+ 17.98\%}$,      1.924 $_{-  0.338}^{+  0.263}$ ) &  (   0.003 $_{-  0.119}^{+  0.117}$   or $_{-3773.88\%}^{+3711.23\%}$,      0.008 $_{-  0.189}^{+  0.192}$ ) \\[4pt]

HD\,103720 &  (  -0.072 $_{-  0.006}^{+  0.006}$   or $_{-  7.72\%}^{+  7.76\%}$,     -0.422 $_{-  0.036}^{+  0.036}$ ) &  (  -0.029 $_{-  0.005}^{+  0.006}$   or $_{- 19.14\%}^{+ 19.25\%}$,     -0.172 $_{-  0.036}^{+  0.036}$ ) \\[4pt]

HD\,200143 &  (  -0.317 $_{-  0.022}^{+  0.024}$   or $_{-  6.79\%}^{+  7.55\%}$,     -6.486 $_{-  0.440}^{+  0.488}$ ) &  (  -0.149 $_{-  0.020}^{+  0.021}$   or $_{- 13.32\%}^{+ 14.13\%}$,     -3.051 $_{-  0.406}^{+  0.431}$ ) \\[4pt]

BD-213253 &  (  -0.169 $_{-  0.037}^{+  0.039}$   or $_{- 22.00\%}^{+ 23.06\%}$,      4.173 $_{-  0.951}^{+  0.908}$ ) &  (  -0.203 $_{-  0.030}^{+  0.032}$   or $_{- 14.72\%}^{+ 15.82\%}$,      4.955 $_{-  0.781}^{+  0.728}$ ) \\[2pt]

\hline  
\noalign{\vskip0.01\columnwidth} 

HD\,216770 &  (   0.014 $_{-  0.006}^{+  0.006}$   or $_{- 44.38\%}^{+ 44.67\%}$,     -0.418 $_{-  0.194}^{+  0.193}$ ) &  (   0.011 $_{-  0.006}^{+  0.006}$   or $_{- 56.18\%}^{+ 56.89\%}$,     -0.336 $_{-  0.193}^{+  0.191}$ ) \\[2pt]

\hline

\end{tabular}
\end{table*} 

The first aspect coming out of Table\,\ref{BISES_simul} is that the variation of $BIS^+$ is larger than that of $BIS^-$ by a factor of 2-3 for all the cases presented. This is associated with stronger correlations and correlation slopes for the former indicator. Moreover, one concludes that the $BIS^+$ leads to a variation that is larger than that of $BIS$ by $\sim$50\,\% for all the configurations when compared it with the corresponding values of the tables in the previous section; the correlations slopes are larger, with the correlation coefficient being very similar. 

The analysis of the results on real data, presented in Table\,\ref{BISES_real} and \ref{BISES_real_MC}, are not easy to interpret. The amplitude variations are larger for $BIS^+$ than for $BIS^-$; the correlation coefficients and correlation slopes are similar or larger for $BIS^+$ than for $BIS^-$ with one clear exception, BD-213253. For the latter, the $BIS^-$ correlation coefficient even approaches the 10\,\% probability level, hinting at the existence of a correlation. This is very interesting in itself, because neither the $BIS$ nor the $V_{span}$ alone could reach this level. However, the high probability of obtaining these correlations by chance prevent us from taking this discussion further.

It is also interesting to compare the results of $BIS$ with those of $BIS^+$ for the real data. The amplitude variation of $BIS^+$ can be larger by a factor of more than 2 than that of the $BIS$ for some cases, but they deliver essentially the same results when it comes to the detection of correlations. Interestingly, an increase in amplitude can even be accompanied by a decrease in the correlation factor; the results on the planet-host star are an interesting example of this. The same happened on BD-213253, decreasing the correlation coefficient to completely negligible levels. This behavior underlines the fact that the variation in amplitude cannot be interpreted straightforwardly. Moreover, it is important to note that that the spectra considered here are high S/N spectra and that the CCF has then a very high S/N ratio; $BIS^+$ is expected to be much less advantageous when the S/N decreases. 

The main conclusion from the analysis of the $BIS^+$ is that it tends to be advantageous for the cases considered here, but it presents a complex behaviour, ranging from clearly advantageous to slightly disadvantageous, and it did not allow to detect any new correlations in the data.


\section{The bi-Gaussian fitting}

\subsection{The method}

\begin{figure}

\includegraphics[width=9cm]{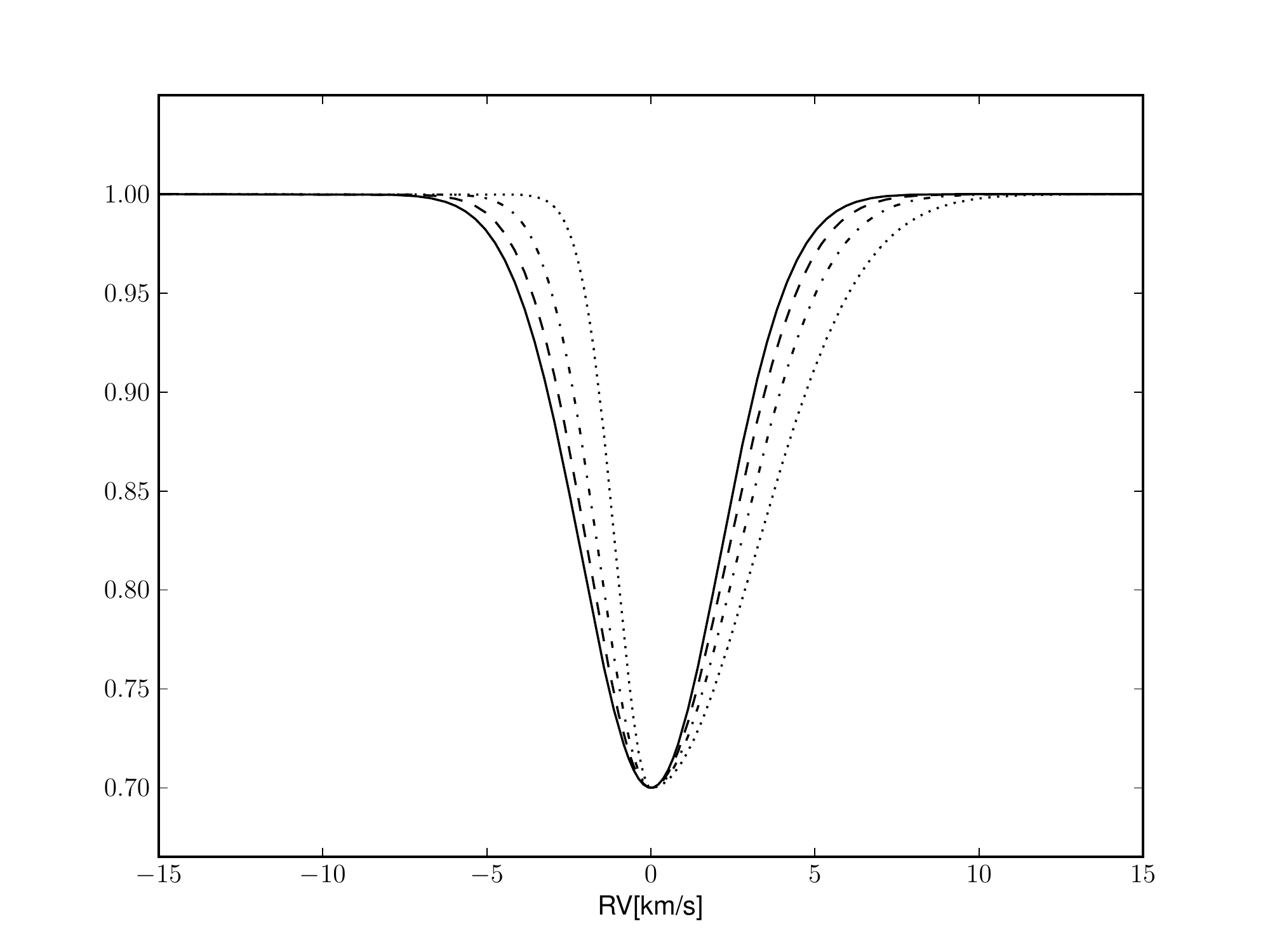}

\caption{Comparison of bi-Gaussian functions with different asymmetry values: 0\% (Gaussian function - {\it solid line}), 10\% ({dashed line}), 25\% ({\it dashed-dotted line}), and 50\% ({\it dotted line}). These normalized functions have a contrast of 30\%, a FWHM of 5\,km/s, and are all centered at 0\,m/s. }\label{BiGaussSketch}

\end{figure}

The bi-Gaussian is a simple function that considers a Gaussian with wings characterized by two different {\it HWHM} (half-width at half-maximum), or, equivalently, an asymmetry parameter associated with the {\it FWHM} (full-width at half-maximum). It was first used in the context of $RV$ measurements by \cite{2006A&A...453..309N} as a way of simultaneously measuring the asymmetry and the center of spectral lines on pulsating stars. It can be represented by

\begin{equation}
bG(\mathrm{RV}) = - D\,\mathrm{exp} \left( - \frac{\mathrm{4\,ln\,2(}RV - RV_{\mathrm{center}})^2}{(FWHM \times (1+A))^2} \right) + C\,\,\, \mathrm{if}\,\, RV > RV_{\mathrm{center}}
\end{equation}

and 

\begin{equation}
bG(\mathrm{RV}) = - D\,\mathrm{exp} \left( - \frac{\mathrm{4\,ln\,2(}RV - RV_{\mathrm{center}})^2}{(FWHM \times (1-A))^2} \right) + C\,\,\, \mathrm{if}\,\, RV < RV_{\mathrm{center}}
\end{equation}

in which {\it D} is the depth of the line, and {\it C} the continuum level, as measured in photoelectrons\footnote{It is equivalent to work with a normalized function, as \cite{2006A&A...453..309N} chose to do.}. Since we apply it to the CCF, we work in the $RV$ space, and both the bi-Gaussian center $RV_{\mathrm{center}}$ and {\it FWHM} are expressed in km/s. The asymmetry {\it A} is given as a percentage (of the {\it FWHM}). For illustration, we plot the effect of introducing an asymmetry on a Gaussian function in Fig.\,\ref{BiGaussSketch}.

This function is fitted by minimizing the $\chi^2_{red}$, just like the Gaussian function; by imposing {\it A}\,=\,0 one recovers the typical Gaussian fitting. We denote the $RV_{\mathrm{center}}$ parameter obtained by Gaussian fitting by $RV_{\mathrm{G, center}}$. As stated by  \cite{2006A&A...453..309N}, the indicator $\Delta V = RV_{\mathrm{G, center}}\,-\,RV_{\mathrm{center}}$ represents the $RV$ shift that can be explained by the asymmetry alone, and we consider it as our proxy for the line deformation.
The interpretation is once again very simple. The extra degree of freedom in the bi-Gaussian fitting allows to accommodate a line profile variation and is thus more insensitive to the line deformations than a simple Gaussian fitting. The operational similarities between $BIS$ and $\Delta V$ then follow very naturally. In this work we considered $\Delta V = RV_{\mathrm{center}}\,-\,RV_{\mathrm{G, center}}$, so that a plot $\Delta V$-$RV$ displays a positive correlation.

\subsection{$\Delta V$: Application to simulated data}\label{Sect_DeltaVsimul}

\begin{figure*}

\includegraphics[width=18cm]{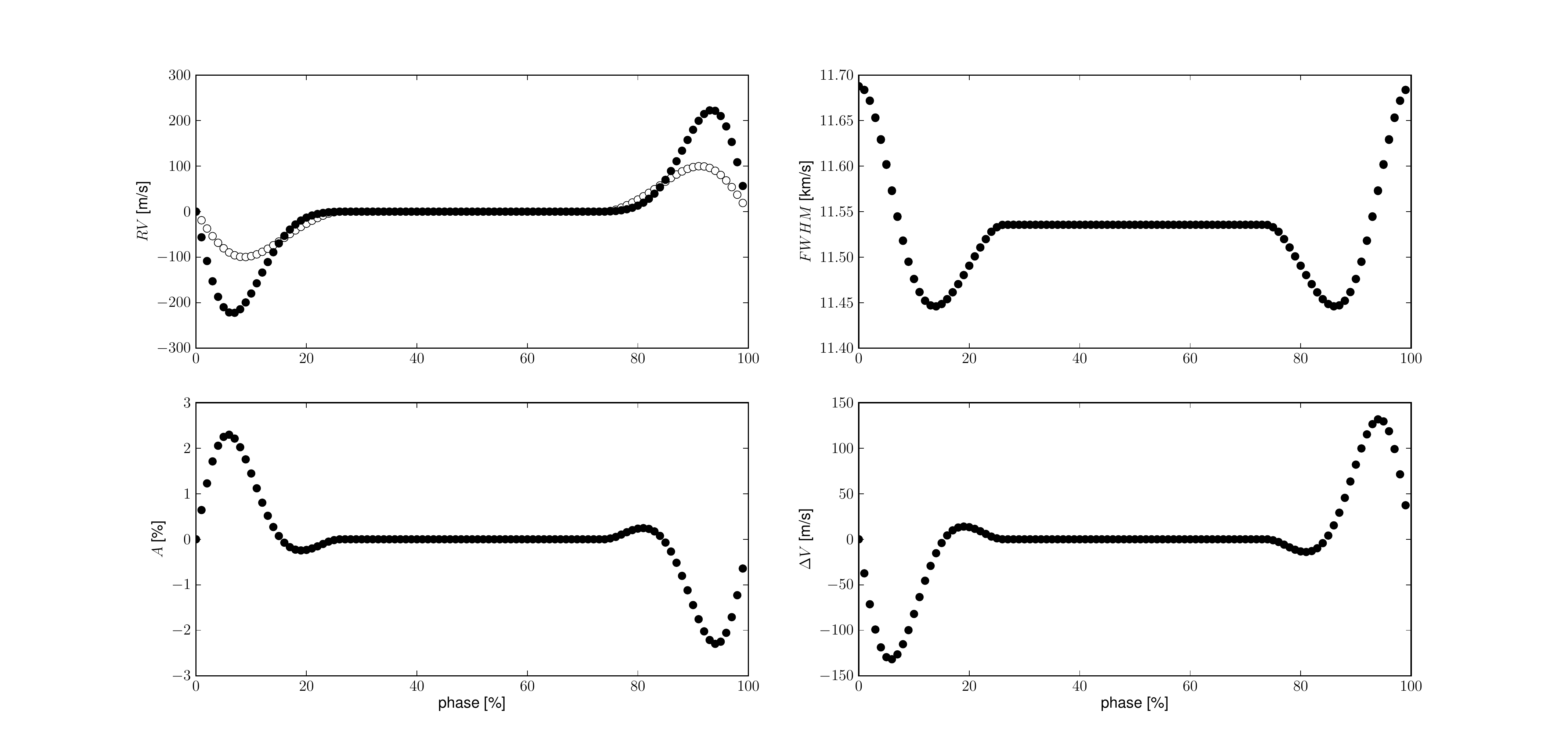}

\caption{Plot of the most important properties for the two fitted functions: $Gaussian$ as open circles (only two upper panels, partially superposed) and $Bi$-$Gaussian$ as filled circles, as a function of orbital phase (with an arbitrary zero-point) for the case of an equatorial spot with a a filling factor of 1\% on an edge-on star with $v\,\sin{i}$ of 7\,km/s. {\it Upper left}: Center of the two fitted functions as a function of orbital phase; {\it Upper right}: $FWHM$ of the two functions; On the two lower panels, we have the asymmetry $A$ ({\it bottom left}) and the $\Delta V$ indicator ({\it bottom right}) for the $Bi$-$Gaussian$ fitting.}\label{example_1}

\end{figure*}

\begin{figure}

\includegraphics[width=9cm]{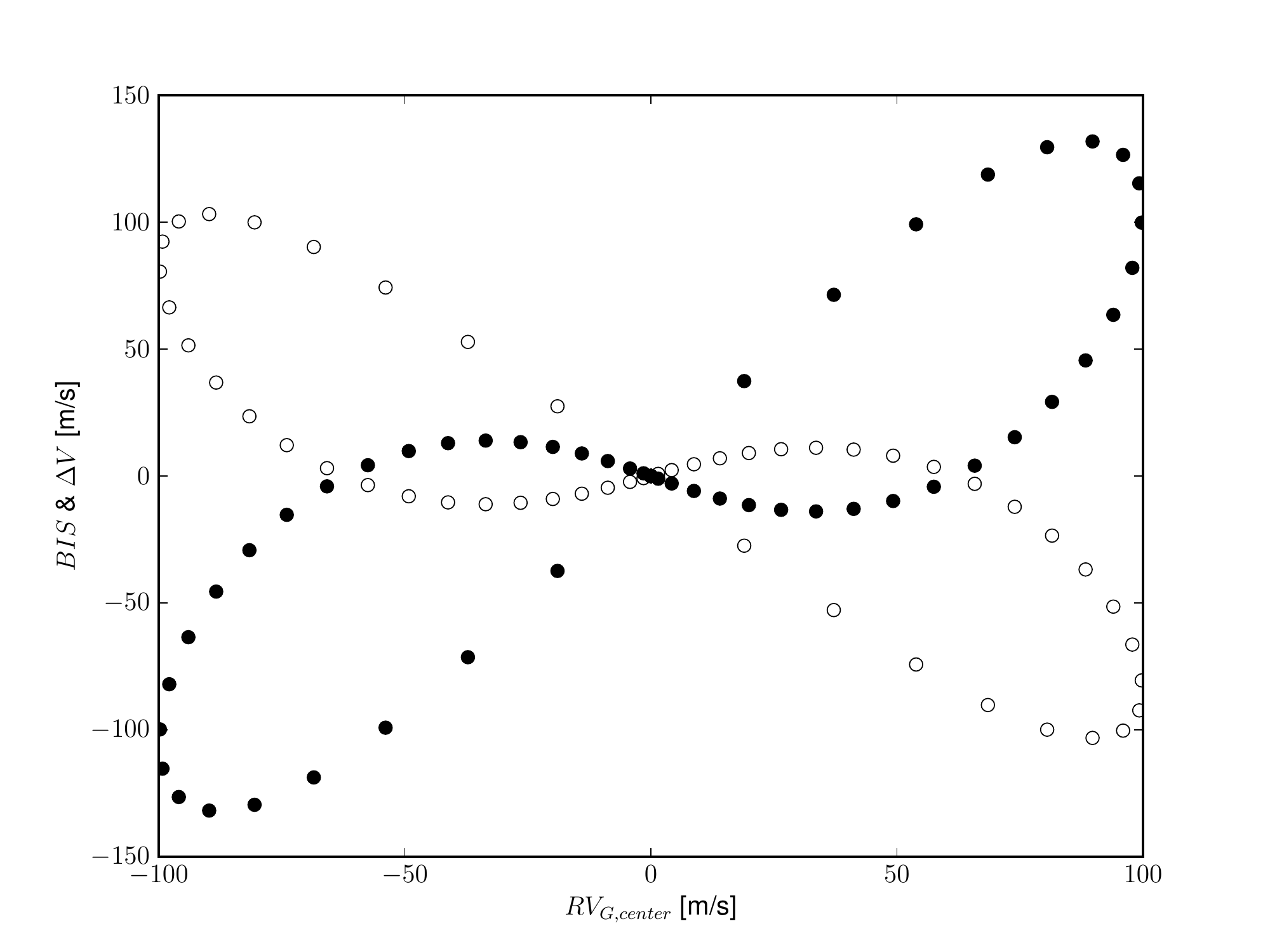}

\caption{Plot of the two indicators considered, $BIS$ (as open circles) and $\Delta V$ (as filled circles), as a function of RV, for the case of an equatorial spot with a filling factor of 1\% on an edge-on star with $v\sin{i}$ of 7\,km/s.}\label{example_2}

\end{figure}

The results for the simulated data are presented in Table\,\ref{simulated}. In Fig.\,\ref{example_1} we plot the phase dependence of the most important parameters for the case of the 1\,\% filling factor and $BIS$ and $\Delta V$ as a function of $RV_{\mathrm{G, center}}$ in Fig.\,\ref{example_2}.

It is easy to notice that it is advantageous to use $\Delta V$ relative to the $BIS$ as an indicator of line-profile deformations for the simulated data. The amplitude of its variation is always higher with the improvement being of 10-30\% (with the most significant improvement being obtained for the largest projected rotational velocity). The value of the correlation slope is also larger, with a variable improvement, depending on the case considered, but reaches above 50\%. These two factors suggest that the $\Delta V$-$RV$ correlation and its slope will emerge more easily in cases where the $BIS$-$RV$ slope is small when compared with the photon noise level. The absolute value of Pearson's correlation coefficient is very high and very similar for both correlations, and the probability that these happen by chance is smaller than 0.01\%. In a nutshell, we can note that the application of the indicator is advantageous when compared with that of $BIS$ for the whole range of spot filling factors and $v.\sin{i}$ considered in the synthetic data, since the amplitude variations are higher with all other parameters being similar. From what we discussed in Section\,4.2 we note that $BIS^+$ variations are higher than those of $\Delta V$ by $\sim$30\%, while displaying similar correlation coefficients.

\begin{table*}

\caption{Comparison between correlations of $BIS$ and $\Delta V$ with $RV$ for simulated data.} \label{simulated}

\centering
\begin{tabular}{cccccccccc} \hline\hline
 \ \ star & $RV_{max}$ & $BIS_{max}$ & $\Delta V_{max}$ & $m_{BIS}$ & $m_{\Delta V}$ & $\rho_{BIS}$ & $\rho_{\Delta V}$ & $\sigma_{\rho, BIS}$ & $\sigma_{\rho, \Delta V}$ \\ 
\hline

HD\,166435' & 9.28e+01	 & 	1.03e+02 	 & 	1.33e+02	 & 	-1.068	 & 	1.364	 & 	-0.990	 & 	0.987 &  9.82 &  9.82 \\

\hline  

$F_r$\,=\,1.0\% & 9.98e+01	 & 	1.03e+02 	 & 	1.32e+02	 & 	-0.686	 & 	0.871	 & 	-0.827	 & 	0.822 &  8.20 &  8.20 \\
$F_r$\,=\,2.0\% & 2.01e+02	 & 	2.10e+02 	 & 	2.66e+02	 & 	-0.679	 & 	0.859	 & 	-0.820	 & 	0.812 &  8.17 &  8.08 \\
$F_r$\,=\,5.0\% & 5.20e+02	 & 	5.50e+02 	 & 	6.94e+02	 & 	-0.655	 & 	0.811	 & 	-0.795	 & 	0.777 &  7.92 &  7.74 \\
$F_r$\,=\,10.0\% & 1.10e+03	 & 	1.22e+03 	 & 	1.47e+03	 & 	-0.599	 & 	0.701	 & 	-0.741	 & 	0.703 &  7.38 &  7.00 \\

\hline 

$v.\sin{i}$=2.0  & 2.21e+01	 & 	1.94e+00 	 & 	2.30e+00	 & 	-0.057	 & 	0.067	 & 	-0.801	 & 	0.802 &  7.98 &  7.96 \\
$v.\sin{i}$=5.0  & 6.43e+01	 & 	3.72e+01 	 & 	4.39e+01	 & 	-0.378	 & 	0.448	 & 	-0.815	 & 	0.817 &  8.10 &  8.15 \\
$v.\sin{i}$=10.0  & 1.57e+02	 & 	2.54e+02 	 & 	3.83e+02	 & 	-1.109	 & 	1.596	 & 	-0.847	 & 	0.817 &  8.46 &  8.12 \\

\hline

\end{tabular}

\tablefoot{When not stated, the probability of obtaining the presented $\rho$ values from the distribution obtained by bootstrapping is lower than 1.0e-4. $RV$, $BIS$, and $\Delta V$ in units of m/s.}

\end{table*}

\subsection{$\Delta V$: Application to HARPS data}\label{Sect_DeltaVreal}

At this point, it is important to recall the discussion on the uncertainties of the estimators, and, in particular, of $BIS$, as presented in Sect.\,3.2. Yet, the error bars on $\Delta V$ are a different issue: These result from the subtraction of the $RV$'s obtained by fitting the same CCF with two different functions. If one assumes that the fitting error is negligible and the final error is dominated by the photon-noise contribution as for $BIS$ when fitting a function to the data, the error bars on the difference are the same as those of either fitting. In other words, the error of both fitting procedures is expected to be completely correlated. To test the two approaches on an even footing, we consider however the error bars on $\Delta V$ as being also $\sqrt{2}$ times the photon noise, just like for $BIS$.

As done before, we separate the results into two tables: The first contains the quantities that do not depend on the error values and their assumptions, as seen in Table\,\ref{real}. The second lists the fitted correlation slopes and uncertainties drawn from bootstrapping the indicators, which depend on the indicator's errors, as seen in Table\,\ref{real_MC}. The reader thus should keep in mind the assumptions behind the results of this last Table.

\begin{table*}

\caption{Comparison between $BIS$ and $\Delta V$ correlations with $RV$ for HARPS data for the parameters that do not consider the error bars.} \label{real}

\centering
\begin{tabular}{cccccccc} \hline\hline
 \ \ star & $RV_{max}$ & $BIS_{max}$ & $\Delta V_{max}$ & $\rho_{BIS}$ & $\rho_{\Delta V}$ & $\sigma_{\rho, BIS}$ & $\sigma_{\rho, \Delta V}$ \\ 
\hline

HD\,224789 &  3.53e+01	 & 	2.77e+01 	 & 	7.73e+01	 &  	-0.818	 & 	0.824 &  4.82 ($<$0.00 \%)  &  4.87 ($<$0.00 \%)  \\

HD\,36051 & 2.16e+01	 & 	3.29e+01 	 & 	7.38e+01	 &  	-0.275	 & 	0.388 &  0.95 (34.31 \%) &  1.34 (18.05 \%) \\

HD\,103720 &  1.10e+02	 & 	2.41e+01 	 & 	6.15e+01	 &  	-0.243	 & 	0.252 &  1.68 ( 9.23 \%) &  1.74 ( 8.17 \%) \\

HD\,200143 &  4.70e+01	 & 	1.78e+02 	 & 	1.80e+02	 &  	-0.302	 & 	0.305 &  1.94 ( 5.30 \%) &  1.96 ( 5.06 \%) \\

BD-213253 &  5.33e+01	 & 	3.33e+01 	 & 	6.16e+01	 &  	-0.130	 & 	0.046 &  0.69 (49.17 \%) &  0.24 (80.69 \%) \\

\hline  

HD\,216770 &  2.67e+01	 & 	4.95e+00 	 & 	1.42e+01	 &  	0.142	 & 	-0.267 & 0.754 (45.10 \%) & 1.408 (15.93 \%) \\

\hline

\end{tabular}

\tablefoot{$RV$, $BIS$, and $\Delta V$ in m/s.}

\end{table*}

\begin{table*}

\caption{Comparison between $BIS$ and $\Delta V$ correlations with $RV$ for HARPS data for the parameters that consider the error bars, i.e. the slope fit parameters.} \label{real_MC}

\centering
\begin{tabular}{ccc} \hline\hline
 \ \ star & $(m,b)_{BIS}$ & $(m,b)_{\Delta V}$  \\
 \hline
\noalign{\vskip0.01\columnwidth}

HD\,224789 &  (  -0.705 $_{-  0.011}^{+  0.012}$   or $_{-  1.57\%}^{+  1.73\%}$,     26.083 $_{-  0.451}^{+  0.408}$ ) &  (   0.831 $_{-  0.013}^{+  0.012}$   or $_{-  1.56\%}^{+  1.40\%}$,    -30.715 $_{-  0.428}^{+  0.479}$ ) \\[4pt]

HD\,36051 & ( -0.476 $_{-  0.125}^{+  0.143}$   or $_{- 26.33\%}^{+ 30.14\%}$,      0.795 $_{-  0.232}^{+  0.202}$ ) &  (   0.635 $_{-  0.153}^{+  0.128}$   or $_{- 24.09\%}^{+ 20.18\%}$,     -1.053 $_{-  0.207}^{+  0.247}$ ) \\[4pt]

HD\,103720 &  (  -0.049 $_{-  0.005}^{+  0.006}$   or $_{- 11.24\%}^{+ 11.35\%}$,     -0.287 $_{-  0.036}^{+  0.036}$ ) &  (   0.061 $_{-  0.006}^{+  0.006}$   or $_{-  9.14\%}^{+  9.04\%}$,      0.367 $_{-  0.037}^{+  0.036}$ ) \\[4pt]

HD\,200143 &  (  -0.239 $_{-  0.021}^{+  0.022}$   or $_{-  8.60\%}^{+  9.27\%}$,     -4.887 $_{-  0.420}^{+  0.453}$ ) &  (   0.259 $_{-  0.023}^{+  0.021}$   or $_{-  8.75\%}^{+  7.98\%}$,      5.303 $_{-  0.462}^{+  0.422}$ ) \\[4pt]

BD-213253 &  (  -0.225 $_{-  0.032}^{+  0.034}$   or $_{- 14.16\%}^{+ 15.25\%}$,      5.527 $_{-  0.838}^{+  0.779}$ ) &  (   0.184 $_{-  0.035}^{+  0.033}$   or $_{- 19.07\%}^{+ 17.90\%}$,     -4.524 $_{-  0.804}^{+  0.857}$ ) \\[2pt]

\hline  
\noalign{\vskip0.01\columnwidth} 

HD\,216770 &  (   0.014 $_{-  0.006}^{+  0.006}$   or $_{- 42.95\%}^{+ 42.81\%}$,     -0.439 $_{-  0.192}^{+  0.193}$ ) &  (  -0.028 $_{-  0.006}^{+  0.006}$   or $_{- 22.15\%}^{+ 22.20\%}$,      0.858 $_{-  0.193}^{+  0.192}$ ) \\[2pt]

\hline

\end{tabular}
\end{table*}

\begin{figure}

\includegraphics[width=9cm]{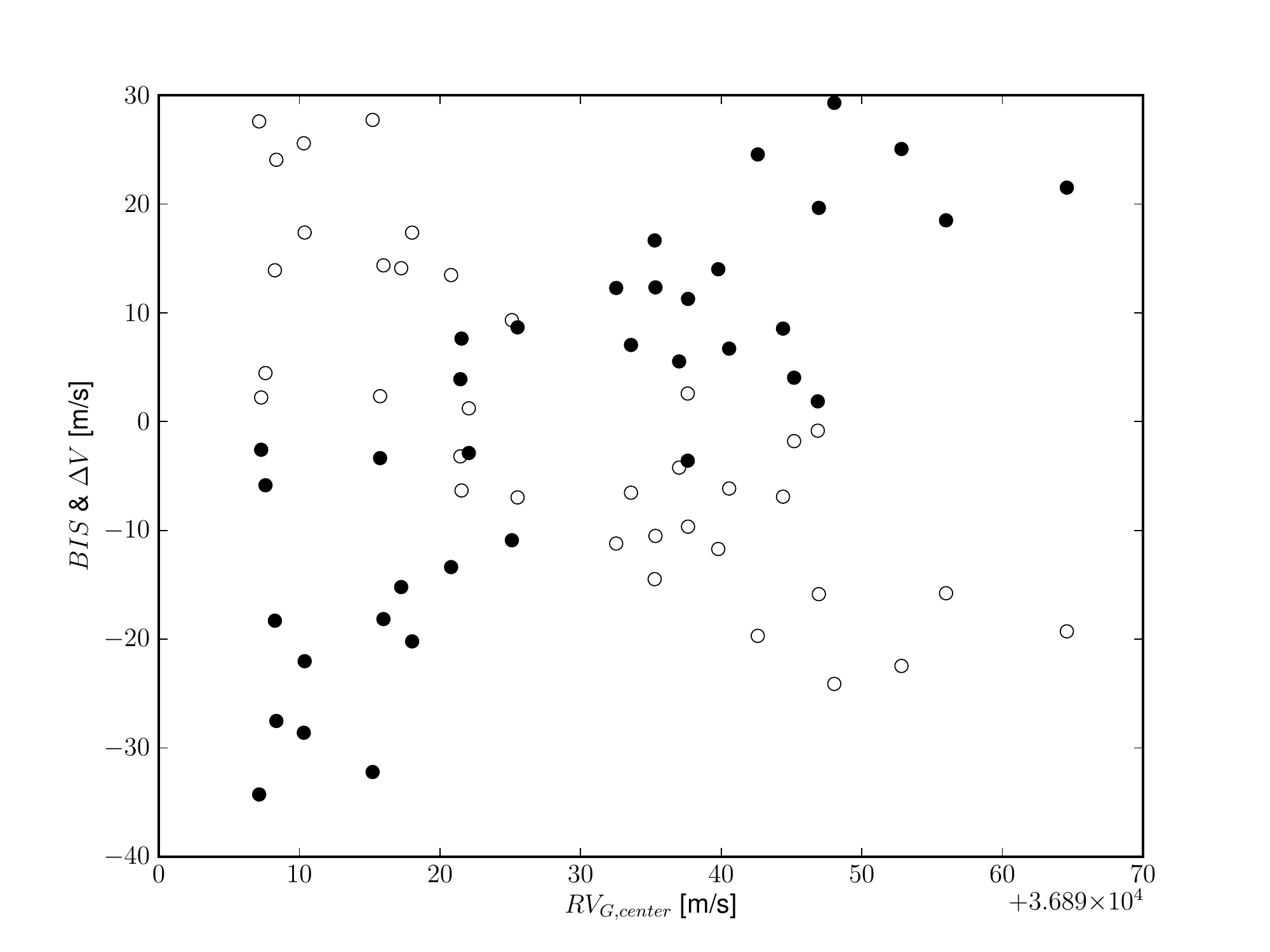}

\caption{Dependence between the indicator $\Delta V$ (filled circles) and $BIS$ (open circles) with $RV$ for HD\,224789. }\label{real_data}

\end{figure}

When applying $\Delta V$ to real data, the situation is slightly different than that of the application to simulated data. The amplitude of $\Delta V$ and the absolute value of Pearsons's correlation coefficient are higher than those of $BIS$, but for both planet-hosting and non-planet hosting stars. Consequently, the probability that the correlation happens by chance tends to be higher with the exception on both of these properties being that of BD-213253. In particular, a more pronounced correlation is obtained for the planet-host star HD\,216770. As expected, the statistical test however shows that one has a non-negligible probability of attaining the measured $\rho$ by chance, which shows that the correlation is not to be trusted, just as for the $BIS$. We plot the application of the indicator to the star with the most significant correlation, HD\,224789, and compare it to $BIS$ in Fig.\,\ref{real_data}. Analyzing the error bars on the slope $m$ is more difficult, and the results are more heterogeneous. As discussed before, these results stand on additional assumptions. The first point to note is that the relative error bars (in \%) for the $\Delta V$ are always smaller than those for the $BIS$, except for the case of the star for which no correlation was detected, BD-213253. As for the others, the absolute values for $m$ are always larger, as expected, but the results are statistically very similar. We also note that all the slopes are non-compatible with zero at several sigma, if the error bars are trusted, even those from BD-213253 and our planet-host, HD\,216770. As a concluding remark on the topic, we note that the results from this indicator are very similar to those obtained using $BIS^+$ on the same stars.

In a nutshell, the $\Delta V$ indicator proves to have a quantitative advantage over the $BIS$ in the sense that the same results are obtained with higher significance, but no qualitative difference in the sense that the same conclusions on each signal's nature were reached.

The $\Delta V$, in spite of a very different construction from that of $BIS$, shares important traits with this other indicator; in particular, both display an 8'-like shape in their correlation with RV (see Fig.\,\ref{example_2}). This suggested that we might be in presence of a common property of profile analysis indicators that have been created so far and it encouraged us to devise a completely different approach, as described in the next section.

\section{The analysis of the asymmetry in the RV calculation: $V_{asy}$}

\subsection{The $V_{asy}$ indicator}

We consider now an indicator that evaluates the RV information content of a spectral line and compares the information on the blue wing (i.e., for wavelengths shorter than that of the line center) to that on the red wing (i.e., for wavelengths longer than the center). For every flux level, the spectral information difference is calculated, and the results for all fluxes are condensed onto a weighted average value for the line in which the weight is the weight attributed to the flux level. The average spectral information between the red and blue wing is used as a proxy for the weight of the undeformed line, i.e., as a weight for the flux level. We apply this procedure to the CCF and mathematically define this indicator, $V_{\mathrm{asy}}$, as

\begin{equation}
V_{asy} = \frac{\sum_{flux} (W_{i}(red) - W_{i}(blue)) \times \overline{W_{i}} }{\sum_{flux} \overline{W_{i}}}\label{Vasy}
\end{equation}

in which $W_{i}$ are the weights of the point calculated at the flux level $i$, as defined in Eq.\,8 of \cite{2001A&A...374..733B}, where

\begin{equation}
W_{i} = \frac{\lambda^2(i)(\partial A_0(i)/\partial \lambda(i))^2 }{ A_0(i) + \sigma_D^2}
\end{equation}

with $\lambda(i)$ and $A_0(i)$ representing the wavelength and flux of the spectra, respectively, and $\sigma_D^2$ the detector readout noise. Since we are working in the velocity space, the wavelength is a velocity. This is standard procedure and has been applied extensively to the CCF. Moreover we are working in a very high S/N domain with the CCF and as such, $A_0(i) + \sigma_D^2 \approx A_0(i)$. We can then rewrite the previous equation in the form of

\begin{equation}
W_{i} = \frac{RV^2(i)(\partial A_0(i)/\partial RV(i))^2 }{ A_0(i)}
\end{equation}

in which $RV(i)$ is the RV value of each pixel (i.e. x-axis) of the CCF (not to be mistaken by the center of the CCF obtained by Gaussian fitting, $RV_{G, center}$).

For every flux level $i$, we consider three quantities: the blue wing information, $W_i(blue)$, the red wing information, $W_i(red)$, and the average between the two, $\overline{W_i} = (W_i(blue) + W_i(red))/2$, as representative of the weight of the information content at this flux level for an undeformed line.

We calculate the flux levels $i$ by quadratic interpolation between the CCF flux levels. We considered 100 equidistant flux levels and excluded the first and last 5 from the sum, to remove the points closer to the continuum and  to the bottom of the line, where the spectral information is small compared with the noise.

In a nutshell, this indicator, $V_{asy}$, is simply the (dimensionless) average information content difference between the left and the right wing of the line. It is much closer to the RV calculation principle than line profile indicators such as $BIS$ or $\Delta V$. Just as for the previously discussed indicators, our indicator's variation is expected to be correlated with that of RV for the case of a signal rooted in a line deformation.

\subsection{$V_{asy}$:Application to simulated data}

\begin{figure}

\includegraphics[width=9cm]{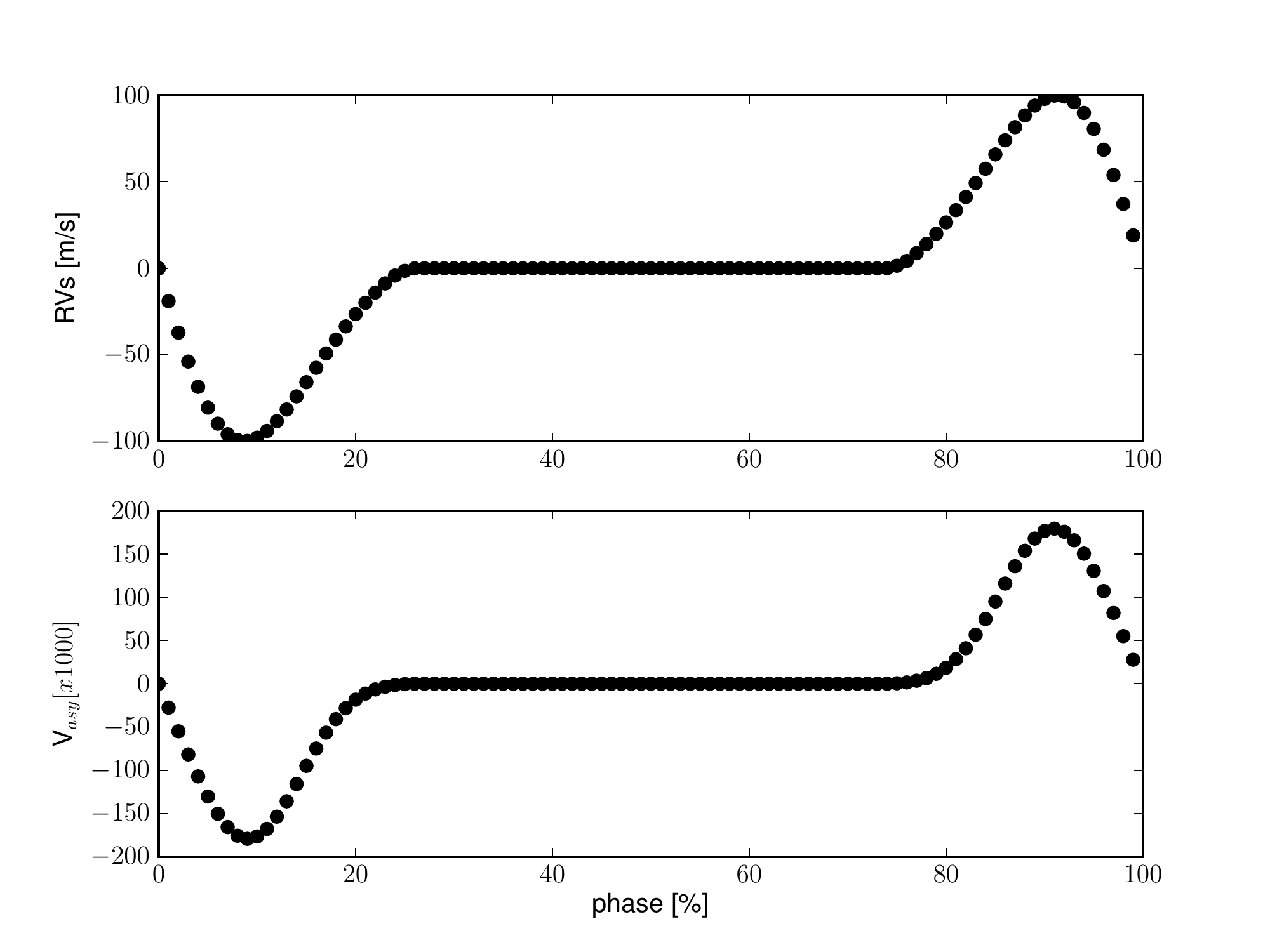}

\caption{Plot of $RV$ ({\it top}) and $V_{asy}$ ({\it bottom}) as a function of orbital phase (with an arbitrary zero point) for the case of an equatorial spot with a a filling factor of 1\% on an edge-on star with $v\sin{i}$ of 7\,km/s.}\label{exampleVasy_1}

\end{figure}

\begin{figure}

\includegraphics[width=9cm]{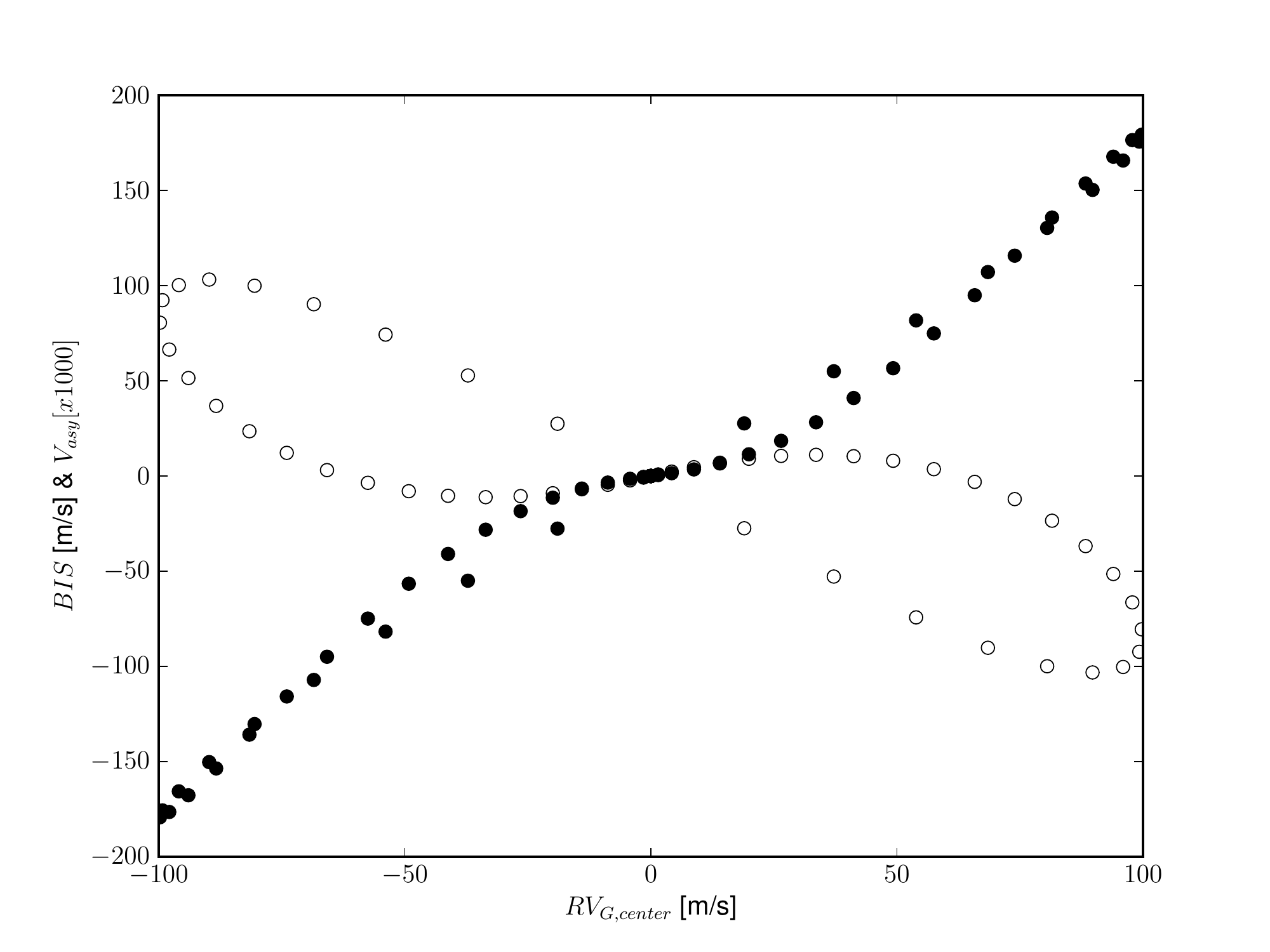}

\caption{Plot of the two indicators considered, $BIS$ (as open circles) and $V_{asy}$ (as filled circles), as a function of RV for the case of an equatorial spot with a a filling factor of 1\% on an edge-on star with $v\sin{i}$ of 7\,km/s.}\label{exampleVasy_2}

\end{figure}

The results of the application of this indicator to the simulated data are presented in Table\,\ref{Vasy_simul}. In Fig.\,\ref{exampleVasy_1}, we plot the variation of $RV$ and $V_{asy}$ as a function of phase for the reference case of an equatorial spot with a filling factor of 1\% on a edge-on star with a $v\,\sin{i}$ of 7\,km/s, and in Fig.\,\ref{exampleVasy_2}, we plot how the correlation between $BIS$ and $RV$ compares to that between $V_{asy}$ and $RV$.

\begin{table*}

\caption{Comparison between correlations of $BIS$ and $V_{asy}$ with $RV$ for simulated data.} \label{Vasy_simul}

\centering
\begin{tabular}{cccccccccc} \hline \hline
 \ \ star & $RV_{max}$ & $BIS_{max}$ & $V_{asy, max}$ & $m_{BIS}$ & $m_{V_{asy}}$ & $\rho_{BIS}$ & $\rho_{V_{asy}}$ & $\sigma_{\rho, BIS}$ & $\sigma_{\rho, V_{asy}}$ \\
 \hline 

HD\,166435' & 9.28e+01	 & 	1.03e+02 	 & 	1.62e+02	 & 	-1.068	 & 	1.704 	 & 	-0.990	 & 	0.999	 &  9.80 &  9.98 \\

\hline  

$F_r$\,=\,1.0\% & 9.98e+01	 & 	1.03e+02 	 & 	1.79e+02	 & 	-0.686	 & 	1.636 	 & 	-0.827	 & 	0.991	 &  8.22 &  9.91 \\
$F_r$\,=\,2.0\% & 2.01e+02	 & 	2.10e+02 	 & 	3.59e+02	 & 	-0.679	 & 	1.620 	 & 	-0.820	 & 	0.991	 &  8.17 &  9.85 \\
$F_r$\,=\,5.0\% & 5.20e+02	 & 	5.50e+02 	 & 	8.94e+02	 & 	-0.655	 & 	1.570 	 & 	-0.795	 & 	0.990	 &  7.90 &  9.87 \\
$F_r$\,=\,10.0\% & 1.10e+03	 & 	1.22e+03 	 & 	1.79e+03	 & 	-0.599	 & 	1.474 	 & 	-0.741	 & 	0.989	 &  7.38 &  9.86 \\

\hline 

$v.\sin{i}$=2.0  & 2.21e+01	 & 	1.94e+00 	 & 	3.14e+01	 & 	-0.057	 & 	1.391 	 & 	-0.801	 & 	0.998	 &  7.95 &  9.96 \\
$v.\sin{i}$=5.0  & 6.43e+01	 & 	3.72e+01 	 & 	1.10e+02	 & 	-0.378	 & 	1.576 	 & 	-0.815	 & 	0.988	 &  8.14 &  9.84 \\
$v.\sin{i}$=10.0  & 1.57e+02	 & 	2.54e+02 	 & 	3.02e+02	 & 	-1.109	 & 	1.603 	 & 	-0.847	 & 	0.958	 &  8.46 &  9.52 \\

\hline

\end{tabular}

\tablefoot{When not stated, the probability of obtaining the presented $\rho$ values from the distribution obtained by bootstrapping is lower than 1.0e-4. 
$RV$ and $BIS$ in m/s.}

\end{table*}

The amplitude variation of the indicator and how it compares with that of $BIS$ are now meaningless, since these indicators are not calculated in the same unit. Of greater importance and insight is that the Pearson's correlation coefficients $\rho$ for $V_{asy}$ are always larger than those for the $BIS$ or $BIS^+$; the improvement is often in the range of 10-20\,\%. It then follows naturally that the offset in $\sigma$ relative to an uncorrelated datasets are very high and that the probabilities of obtaining these correlation coefficient values by chance are very low (even though one has to stress that they are negligibly low for all cases considered, as before). 

An inspection of Fig.\,\ref{exampleVasy_2} shows that our objective was accomplished in the sense that the relation between $V_{asy}$ and $RV$ is much closer to a linear one than that with the $BIS$. There is still a departure from linearity for small RV variations, but the usefulness of the method is clear.

\subsection{$V_{asy}$:Application to HARPS data}

\begin{figure}

\includegraphics[width=9cm]{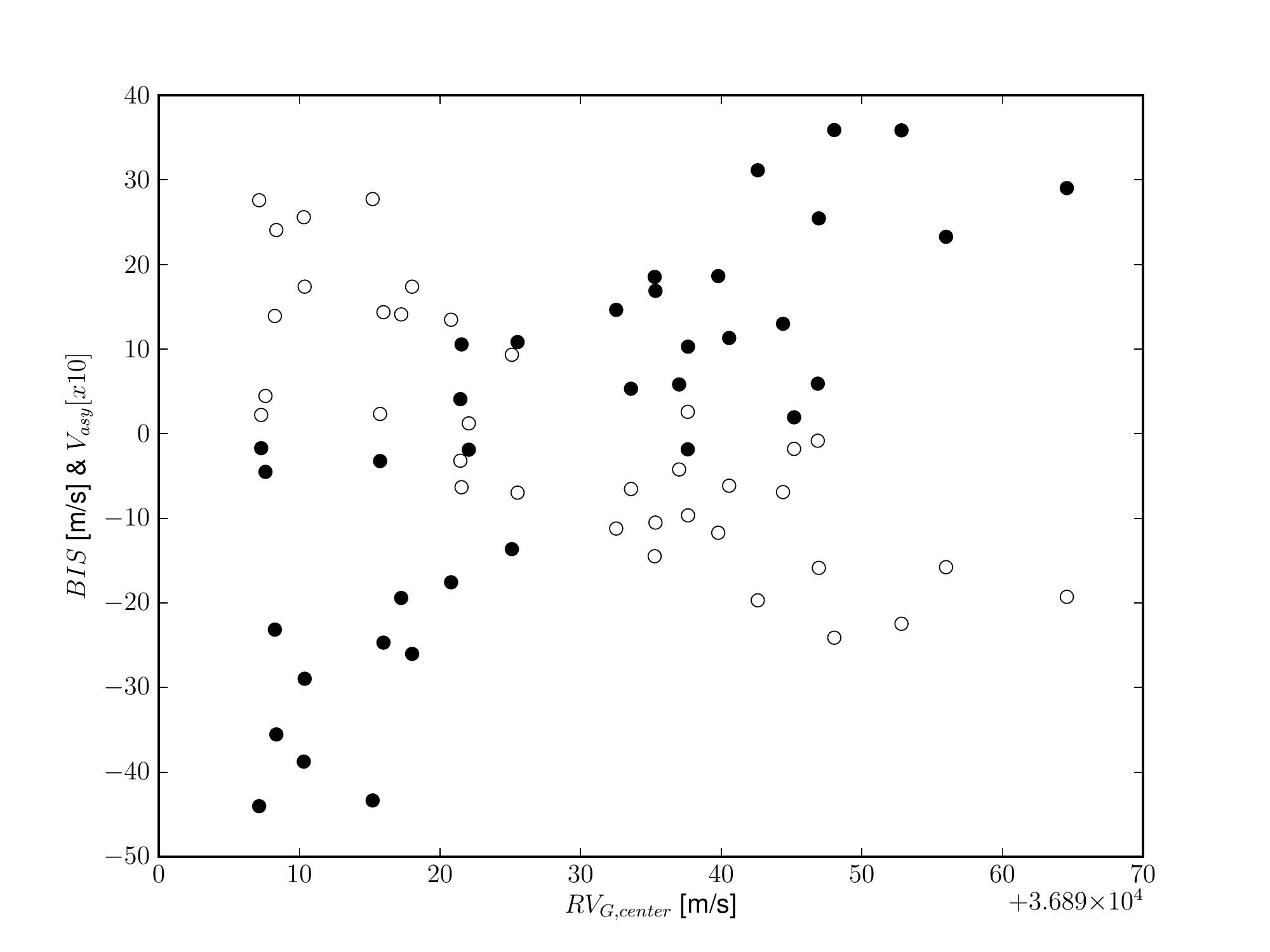}

\caption{Dependence between the indicator $V_{asy}$ (filled circles) and $BIS$ (open circles) with $RV$ for HD\,224789. }\label{Vasy_real_data}

\end{figure}

In Sect.\,\ref{Sect_DeltaVreal}, we applied $\Delta V$ to real data and we now repeat the same procedure and analysis for $V_{asy}$. The only difference is that we refrain from attempting to calculate the error bars on the indicator. As we saw in the mentioned Section, this was the weakest point of the analysis, requiring more assumptions and making the interpretation of results more difficult. As a consequence, we also chose not to calculate the error bars on the correlation slope $m$. The results are presented in Table\,\ref{Vasy_real} and in Fig.\,\ref{Vasy_real_data}, we depict the correlation between $V_{asy}$ and $RV$ and how it compares with that of $BIS$ for the case of HD\,224789.

\begin{table*}

\caption{Comparison between correlations of $BIS$ and $V_{asy}$ with $RV$ for HARPS data.} \label{Vasy_real}

\centering
\begin{tabular}{cccccccccc} \hline \hline
 \ \ star & $RV_{max}$ & $BIS_{max}$ & $V_{asy, max}$ & $m_{BIS}$ & $m_{V_{asy}}$ & $\rho_{BIS}$ & $\rho_{V_{asy}}$ & $\sigma_{\rho, BIS}$ & $\sigma_{\rho, V_{asy}}$ \\ 

\hline

HD\,224789 &  3.53e+01	 & 	2.77e+01 	 & 	4.40e+03	 & 	-0.756	 & 	116.351 	 & 	-0.818	 & 	0.823	 &  4.83 ($<$0.00 \%) &  4.86 ($<$0.00 \%) \\
HD\,36051 & 2.16e+01	 & 	3.29e+01 	 & 	1.25e+02	 & 	-0.376	 & 	3.439 	 & 	-0.275	 & 	0.690	 &  0.95 (34.07 \%) &  2.40 ( 1.66 \%) \\
HD\,103720 & 1.10e+02	 & 	2.41e+01 	 & 	2.88e+02	 & 	-0.043	 & 	1.409 	 & 	-0.243	 & 	0.688	 &  1.69 ( 9.11 \%) &  4.77 ($<$0.00 \%) \\
HD\,200143 & 4.70e+01	 & 	1.78e+02 	 & 	5.19e+04	 & 	-0.443	 & 	59.314 	 & 	-0.302	 & 	0.153	 &  1.94 ( 5.30 \%) &  0.98 (32.64 \%) \\
BD-213153 & 5.33e+01	 & 	3.33e+01 	 & 	1.59e+03	 & 	-0.103	 & 	9.432 	 & 	-0.130	 & 	0.257	 &  0.69 (48.93 \%) &  1.35 (17.54 \%) \\
\hline  

HD\,216770 &  2.67e+01	 & 	4.95e+00 	 & 	8.86e+02	 & 	0.018	 & 	-0.013 	 & 	0.142	 & 	-0.001	 &  0.75 (45.18 \%) &  0.00 (99.68 \%) \\

\hline

\end{tabular}

\tablefoot{$RV$ and $BIS$ in m/s.}

\end{table*}

As before, we skip the analysis of the variation of the coefficient, and we note that while the values of the correlation slopes are higher this does not mean the method is more efficient in itself. These properties are in line with what was measured for the synthetic data, but they do not allow us to reach any conclusion in themselves. However, it is interesting to note that for 4 of the 5 active stars the Pearson's correlation coefficient was significantly improved, in one case the improvement was by a factor of two, and in the case of the planet-hosting star, it decreased. Consequently, the deviation in $\sigma$ increased for the four active stars and the probability of the correlation arising from a fortuitous alignment of the data decreased. The opposite was obtained for the planet-host star. This provides a strong argument for the usefulness of the indicator and illustrates very well how advantageous it can be when compared with the $BIS$.


\section{Discussion}

We have shown that the $\Delta V$ indicator provides a more efficient characterization of line deformations than the $BIS$, and very similar to that of $BIS^+$. This property can make the difference when the correlation slope is comparable to the photon noise and/or the number of observations is very small. The indicator proved to be advantageous over the standard $BIS$ for all the simulated datasets and slightly so for the real ones for which a correlation had been found between the $RV$ and the $BIS$. In this sense, we can then conclude that the $\Delta V$ provides a quantitative advantage over the $BIS$. Yet, it is important to note that some of the active stars show $RV$ variations without being correlated with $BIS$ or $BIS^+$, and our new indicator did not allow to identify new correlations (for the meager number of stars tested). 

Furthermore, the error bars on $\Delta V$ and $BIS$ can be used to assess the significance of the results in another way. If one assumes that the activity signal is the only signal present in the $RV$ data, the residuals should follow the noise distribution of the indicator once the slope is subtracted. This is expected to be a Gaussian distribution with a center at zero and a scatter dictated by the photon noise or, to be more generic, proportional to it. We consider different multiplicative factors $\alpha$ for the relation between the observed scatter and the one dictated by photon noise $\sigma_{obs}=\alpha \times \sigma_{ph}$. We evaluate them through the Kolmogorov-Smirnoff analysis \citep[as implemented in][]{1992nrfa.book.....P}, and evaluate which of these postulated distributions $G(\alpha \, \sigma_{ph})$ has the highest probability of coming from the same parent distribution as the observed one, and thus which $\alpha$ is more appropriate to characterize $\sigma_{obs}$. We considered $\alpha$ over an interval ranging from 0 to 30 with a step of 0.5. We list the highest-probability $\alpha$ obtained for both correlations in Table\,\ref{alpha_corr}.

\begin{table}

\caption{The $\alpha$ correction factor on photon noise used to obtain the observed noise distribution.} \label{alpha_corr}

\centering
\begin{tabular}{lcc} \hline \hline
 \ \ star   &  $\alpha(BIS)$ & $\alpha(\Delta V)$ \\ 
\hline 
HD\,224789 & 14.5  & 20.0 \\
HD\,36051  & 5.5  & 7.0 \\
HD\,103720 & 7.5  & 6.0 \\
HD\,200143 & 7.5  & 5.5\tablefootmark{a} \\
BD-213153 & 5.0  & 5.5 \\

\hline
\end{tabular}

\tablefoottext{a}{Note that the probability of correspondence for $\Delta V$ is smaller than 20\%.}

\end{table}

We can see from the Table that neither of the methods/correlations is intrinsically superior when it comes to noise distribution. In other words, the correction factors applied to the error bars on the indicators are larger for the $BIS$-$RV$ correlation for some of the cases, and larger for the $\Delta V$-$RV$ correlations for some other cases. They are significantly higher than unity in both situations, which should send a warning to all those who use the $BIS$ indicator routinely. Yet, one has to bear in mind that the fit of a straight line on the data is ill-justified, since we know that the real shape is closer to an 8 rather than to a straight line. While the results still hold for a relative comparison, an $\alpha$ larger than unity was always expected. 

We stress again that non-parametric tests are particularly important when the properties of the distributions are unknown or ill-defined, and we recall the usefulness of the permutation test to the reader as a way of assessing the significance of a correlation.

As discussed by \cite{2006A&A...453..309N}, the fitting makes use of the full information of the line and will still deliver precise results at low S/N, which is a level at which $BIS$ is significantly more vulnerable. The reason for this is two-fold: $BIS$ requires interpolation to calculate the flux levels and does not make use of all the spectral information content of the line. It is calculated only over a restricted range of flux. This is particularly important for those who use CCFs constructed from a relatively low number of lines or aim at measuring $RV$ from low-S/N spectra. Neither is common for typical RV planet searches, but the latter is frequent when using RV measurements to confirm transit candidates from space observatories, such as CoRot \citep{2011A&A...531A..41C} or {\it Kepler} \citep{2011ApJS..197...13E, 2012A&A...545A..76S}. Interestingly, a significant source of false positives in transit searches is the blending of the spectral lines with those of a foreground or background star. The usage of indicators, such as the BIS or those presented in this paper, can again be of help, but this is a complex issue and beyond the reach of this paper. We will discuss it in a separate paper, using meaningful false positive examples from transit searches (Santerne et al. {\it in prep}).

The results on $V_{asy}$ are even more promising than those of $\Delta V$, with the indicator being constructed in a completely different way. The results point for more secure correlations between $V_{asy}$ and the RV for most of the analyzed active stars (4 out of 5) compared to those obtained using the $BIS$. It provides both a qualitative and quantitative advantage over the bisector in the sense that new correlations arise for active stars, and those previously identified are detected with higher significance. Yet, it has to be noted that even though the correlation is much closer to a linear one than that of the $BIS$, a departure from the linear relationship is apparent for the smallest amplitude $RV$ variations. This attests for an undersensitivity of the indicator in this regime and warrants further investigation.     

In this work we made a distinction and a separated analysis of synthetic spectra, without noise, and real spectra, with noise. This was a choice prompted by the availability of the SOAP model, which delivers noise-free CCFs, but allowed us to explore and distinguish expected theoretical results and observed ones. While one could have attempted to introduce noise in the CCFs and compare it with the real data, this would have to be done {\it before} the CCF computation, i.e., at the spectral generation level, using a completely different and by far more complicated procedure, which is highly dependent on the particular assumptions on the spectra. By construction, the calculation of the CCF transforms white noise into correlated noise, simply because the correlation step for the calculation is smaller than the pixel size. Moreover, a real spectra is expected to have its own noise correlated from one pixel to the other; this is a consequence of atmospheric extinction being a function of wavelength, just like instrumental transmission and instrumental profile. As a consequence, any model of the spectra which does not fully characterize the spectrograph, would fall short in describing its correlated noise, and thus lead to an incorrect evaluation of its impact on the $RV$ and associated indicator's calculation.

It has been suggested recently that there is a significant difference between the results derived from a simple spot model (such as SOAP) and comprehensive $RV$ campaigns \citep[e.g.][]{2012ApJ...761..164C}. In this work, the parametrized synthetic cases show similar order-of-magnitude variations in $RV$ and indicators compared to the real ones. Yet, one cannot exclude that there might be physical mechanisms that are not represented and might produce an appreciable effect. Works such as those of \cite{2010A&A...519A..66M} showed the importance of considering the reduction of convective blueshift at the spot's location on the measurement of precise stellar $RV$. This is only one of the many points not at all adressed by SOAP, since it emulates a very simple stellar model, and does not even consider emission of any kind at the spot's location, nor the nuances in its emitted spectra. While a cold-spot model might be indeed an oversimplification, we note that the methods discussed here proved to be effective not only for simulated data but for real data too.

It is also important to stress that the RV center obtained for a deformed line depends on the RV calculation method. For the implications of this characteristic on the case of Rossiter-MacLaughlin measurements, the interested reader is referred to \cite{2012arXiv1211.3310B}. This is a point to keep in mind when comparing the results obtained here with those obtained for RVs acquired with the iodine cell technique \citep[e.g.][]{2011ApJ...733..116B}.

We conclude this discussion on a different note: Instead of considering these different methods as alternative options, one should look at them as independent ways of characterizing the same phenomena and use them simultaneously, while bearing in mind their relative strengths and weaknesses. The active star BD-123153 illustrates this very well: It exhibits a non-significant correlation between $BIS$ and $RV$ or $\Delta V$ and RV but shows a more significant one between $V_{asy}$ and $RV$ or $BIS^-$ and $RV$. The use of different tests is then very important, especially because while the presence of a correlation disproves the planetary nature of a signal, its absence does not prove its planetary nature. We tried to do a systematic assessment of the potential of the two presented methods, but the work presented here should be seen merely as the first step toward a more detailed characterization of these diagnosis methods and the development of complementary approaches.

\section{Conclusion}

In this paper, we introduced two diagnosis methods for detecting line-profile variations in the context of exoplanet searches. The objective is to evaluate if these variations can be pinpointed as being at the root of detected $RV$ signals. We attempted at a systematic assessment of the efficiency and sensitivity of our presented methods when compared with that of BIS, assuming different parametrizations, and $V_{span}$, and following very clear criteria. Our conclusions can be outlined as follows:

\begin{itemize}
 \item The standard $BIS$ delivers very similar results to those of $V_{span}$ for the cases considered. Different parametrizations of $BIS$ can be explored to increase the leverage of the diagnosis at the expense of considering a lower fraction of the line and thus working at lower S/N. Yet, statisticaly more significant results can be obtained, showing that this is an option to consider.
 \item The bi-Gauss/$\Delta V$ method is superior to the standard $BIS$ in a multitude of cases. It presents no qualitative advantage, identifying the same signals than $BIS$ as activity-rooted, but provides more significant results with the characterization parameters being affected by smaller error bars. This point can make the difference for cases where the correlation slope between the indicator and $RV$ is close to the noise level and puts the indicator in even footing with the more effective $BIS^+$ parametrization. 
 \item The $V_{asy}$ indicator evaluates the unbalance between the red and blue wings of a line in terms of spectral information, being very different by construction from the other profile indicators. It shows a  significant advantage over different parametrizations of $BIS$ and $\Delta V$, when applied to simulated spectra and to real spectra. The indicator leads to more significant correlations, and identifies new signals rooted in activity, providing a clear qualitative advantage. It also gives rise to a correlation with RV which is much closer to a straight line than any of the other profile-based indicators.
\end{itemize}

Finally we argue for the usage of independent techniques to characterize stellar line profile variations and to identify their origins, creating a coherent view of the stellar phenomena. This is particularly important in cases where the signal and/or potential correlation slopes approach the RV uncertainty level. A more detailed knowledge of the diagnosis methods themselves is required to meet the exigent requirements of today's scientific programs. The work presented here should be considered as a first step in this direction; a more significant amount of resources should be put into characterizing these indicators and devising new and independent ones.


\begin{acknowledgements}
This work was supported by the European Research Council/European Community under the FP7 through Starting Grant agreement number 239953, and by Funda\c{c}\~ao para a Ci\^encia e a Tecnologia (FCT) in the form of grant reference PTDC/CTE-AST/098528/2008 and PTDC/CTE-AST/098604/2008. NCS would further like to thank FCT through program Ci\^encia\,2007 funded by FCT/MCTES (Portugal) and POPH/FSE (EC). We thank the anonymous referee for insightful comments that led to a significant improvement of the overall paper's quality. PF would also like to thank Alexandre Santerne and Mahmoudreza Oshagh for enlightening discussions.

\end{acknowledgements}

\bibliographystyle{aa} 
\bibliography{Mybibliog} 

\end{document}